\documentclass[12pt]{article}
\usepackage{amsmath,amssymb}
\usepackage{graphicx,color}
\numberwithin{equation}{section}
\usepackage{cite}
\usepackage{bm}
\usepackage{dcolumn}
\newcommand{\be}{\begin{equation}}
\newcommand{\bea}{\begin{eqnarray}}
\newcommand{\eea}{\end{eqnarray}}
\newcommand{\ba}{\begin{array}}
\newcommand{\ea}{\end{array}}
\newcommand{\ee}{\end{equation}}

\expandafter\ifx\csname mathbbm\endcsname\relax

\else

\fi \textheight 22cm \textwidth 15cm \topmargin 1mm
\begin{document}
\begin{titlepage}
\hfill
\vbox{
    \halign{#\hfil         \cr
           IPM/P-2008/061 \cr
                      } 
      }  
\vspace*{20mm}
\begin{center}
{\Large {\bf Non-relativistic CFT and Semi-classical Strings }\\
}

\vspace*{15mm}
\vspace*{1mm}
{Amin Akhavan $^{a}$
, Mohsen Alishahiha$^{b}$, Ali Davody$^{a,b}$ and  
Ali Vahedi$^{a,b}$}

 \vspace*{1cm}
{\it ${}^a$ Department of Physics, Sharif University of Technology \\
P.O. Box 11365-9161, Tehran, Iran}

\vspace*{.4cm}

{\it ${}^b$ School of physics, Institute for Research in Fundamental Sciences (IPM)\\
P.O. Box 19395-5531, Tehran, Iran \\ }

\vspace*{.4cm}
{  amin\_akhavan@mehr.sharif.ir, alishah ,davody, vahedi @ipm.ir} 

\vspace*{2cm}
\end{center}

\begin{abstract}
We study different features of 3D non-relativistic CFT using gravity description. 
As the corresponding gravity solution can be 
embedded into the  type IIB string theory, we study semi-classical closed/open strings in this background.  
In particular we consider folded rotating and circular pulsating closed strings where we find
the anomalous dimension of the dual operators as a function of their quantum numbers.
We also consider moving open strings in this background which can be used
to compute the drag force. In particular we find that for slowly moving particles, the
energy is lost exponentially and the characteristic time is given in terms of the temperature, 
while for fast moving particles  the energy loss goes as inverse of the time and the
characteristic time is independent of the temperature.

\end{abstract}

\vspace{2cm}
 \begin{center}
 \it{Dedicated to Reza Mansouri on the occasion of his 60th birthday}
 \end{center}

\end{titlepage}

\section{Introduction}

AdS/CFT correspondence \cite{Maldacena:1997re} has given us a useful tool to find  weakly
coupled descriptions for strongly coupled conformal field theories. The weakly coupled
description generically is given in terms of a supergravity on a background containing an AdS part. 
Actually there is a one to one correspondence between objects in gravity side and those in the conformal 
field theory dual. In particular it is known that the symmetries of the conformal field theory can be
geometrically realized in the gravity side as the isometries of the metric. 
For example the conformal group of the
$d$-dimensional space-time, $SO(d,2)$, is realized as the isometry of the $AdS_{d+1}$ geometry where the
gravity is defined.

Encourage with the great success in describing relativistic strongly coupled conformal
field theories, it is natural to look for a weakly coupled gravity descriptions for 
non-relativistic conformal field theories. Indeed there are several models, for example, in 
condensed mater where the theories are invariant under Schr\"odinger group which is essentially
the conformal group of the non-relativistic field 
theories (for example see \cite{Nishida}). Therefore it is important to find the corresponding 
gravity dual.

Actually taking into account that the Schr\"odinger group, $Sch(d-1)$, is a subgroup of the
relativistic conformal group $SO(d,2)$ and  indeed can  be obtained from it by taking non-relativistic
limit (contraction), one would expect that the same procedure can be applied in the gravity side as well.
Namely one expects that the geometry we are interested in could be given by a deformation 
of AdS geometry. In fact the corresponding gravity whose isometry is $Sch(d-1)$ has been proposed in \cite{{Son:2008ye},{Balasubramanian:2008dm}}\footnote{For recent studies on AdS/non-relativistic
field theories see \cite{{Goldberger:2008vg},{Wen:2008hi},{Nakayama:2008qm},{Chen:2008ad},
{Minic:2008xa},{Imeroni:2008cr},{Galajinsky:2008ig},{Kachru:2008yh},
{Pal:2008rf},{SekharPal:2008uy},{Pal:2008id},{Duval:2008jg},{Lin:2008pi},
{Hartnoll:2008rs},{Schvellinger:2008bf},{Rangamani}} (See also \cite{Leiva:2003kd}).
} which is
\bea 
ds^2=\frac{r^2}{R^2}(2dtdy+d{x}_i^2-\mu^2 r^2dt^2)+\frac{R^2}{r^2}dr^2,
\label{back}
\eea
where $i=1,2,\cdots,d-1$. Here $\mu$ is a parameter which controls the deviation from
AdS geometry. In other words $\mu$ is a parameter which characterizes the non-relativistic nature
of the dual field theory. More precisely, as we will see, the corresponding physical dimensionless parameter 
indeed $\mu R$.
The above metric is invariant under the following rescaling of the coordinates
\bea
 t\rightarrow \lambda^2 t,\;\;\;\;\;x_i\rightarrow \lambda x_i,\;\;\;\;\;y\rightarrow y,\;\;\;\;\;
 r\rightarrow \lambda^{-1} r. 
\eea
This $d+2$-dimensional gravity is proposed to describe a $d$-dimensional non-relativistic
CFT which is invariant under the following scaling
\bea
 t\rightarrow \lambda^2 t,\;\;\;\;\;\;\;\;x_i\rightarrow \lambda x_i,\;\;\;{\rm for}\;\;\;i=1,\cdots,d-1. 
\eea

Since our main knowledge of AdS/CFT duality has come from string theory, it is natural to pose 
the question  whether this solution can be embedded in the ten 
dimensional superstring theories. If it does, then one may use our experiences in string theory
to study some features of the non-relativistic CFTs.

The aim of this article is to study some features of (1+2)-dimensional non-relativistic CFT 
using supergravity solution in type IIB string theory.
Since we are dealing with string theory, it is natural 
to consider a semi-classical string in this background. 
Having this in our mind, we will first consider a folded rotating closed string \cite{Gubser:2002tv} (for more details see, e.g. \cite{Frolov:2002av})
and evaluate the relation between 
its energy, $E$, and spin $S$. Using AdS/CFT dictionary this
corresponds to an operator with anomalous dimension $\Delta=E$ and spin $S$. We find that, for $\mu R\ll1$ 
the anomalous dimension gets logarithmic corrections similar to AdS case, though the coefficients are functions of 
$\mu R$, at least up to order we are considering. This might be interpreted as the fact that in going from 
relativistic to non-relativistic theory the anomalous dimension of the corresponding operators gets
corrections. We will also
study circular pulsating strings following Ref. \cite{Minahan}. In this case, in comparison with 
the AdS case, we find new behavior at subleading order
 which we would like to associate with the non-relativistic
properties of the dual theory.

It is also interesting to study semi-classical open strings in this background. 
In the context of AdS/CFT correspondence
an open string can be associated to a Wilson loop in the dual field theory, which in turn can be used
to compute the effective potential between external objects (e.g. quark-anti quark potential). In our case 
at zero temperature we find the potential behaves as $l^{-2}$ as expected for non-relativistic
CFT with dynamical scale $z=2$. 

To explore non-relativistic properties of the dual field theory
we will also consider a moving open string on the relevant geometry at finite temperature which may be
thought of as a moving {\it external object} on the {\it hot plasma}\footnote{When the dual theory 
is gauge theory the external object could be quark and the plasma can be made out of quark-gluon plasma.}.
In this case, unlike the relativistic case, we observe that there is no upper bound on the velocity of the
moving object, as expected for a non-relativistic theory. We will also see that when the external 
object moves through the plasma very slowly it loses its energy exponentially as a function of time, 
while when it moves very fast the decay rate behaves as inverse of time. 

We will also redo the drag force-like computations for the non-relativistic CFT at zero temperature. 
Being at zero temperature the moving objects still lose their energy, though unlike the finite temperature
case, the characteristic time is given in terms of $\mu$.

The paper is organized as follows. In the next section we review the embedding of the relevant geometry
in type IIB string theory where we will also present the general features of the dual non-relativistic
CFT. In section three we will study semi-classical closed strings on the background presented in 
section two. In section four we consider moving open strings on the geometry generated by a
black hole in the supergravity solution of section two. In the section five 
we study the drag force in the non-relativistic field theory at 
zero temperature. The last section is devoted to discussions and conclusions.

\section{Supergravity description}

The supergravity solution we are interested in can 
be obtained from  the $AdS_5\times S^5$ solution of type IIB supergravity solution  via 
TsT transformation or by making use of the Null Melvin 
Twist procedure \cite{Gimon:2003xk}, \cite{{Alishahiha:2003ru},
{Herzog:2008wg},{Adams:2008wt},{Maldacena:2008wh},
{Mazzucato:2008tr}}\footnote{ Note that the
whole solution beside the metric contains a non-zero RR 4-form as well as non-zero NSNS B field whose
explicit forms are not important for what follows. The dilaton is constant as expected.}
\bea  
ds^2=\left(\frac{r}{R}\right)^2\left[-{\mu^2}{r^2}dt^2+2dtd\xi+dx_i^2\right]
+\left(\frac{R}{r}\right)^2\left[dr^2+r^2d{\cal M}_5^2\right], 
\label{sol1}
\eea
where $d{\cal M}_5^2$ is the metric of the five dimensional internal space whose spin structure
fixes the number of supersymmetries preserved by the background. For example the internal
space could be a five sphere. 
Indeed this solution has been obtained in
\cite{Alishahiha:2003ru} in the context of light like dipole field theory
where it was shown that the solution preserves eight supercharges. 

To proceed let us consider the dilaton field,$\phi$ which can be treated as a massless scalar field 
in the bulk supergravity. In case of $AdS_5\times S^5$, the dilaton field 
is dual to the operator ${\cal O}={\rm Tr} F^2$ with $\Delta=4$ whose two point function is
\be
\langle\; {\cal O}(x,t) {\cal O}(0)\;\rangle\sim \frac{1}{|X|^8},\;\;\;\;\;{\rm with}\;\;|X|=\sqrt{-t^2+x^2},
\ee

In our case we would like to study the dilaton field in the background (\ref{sol1}).
Setting $\phi=e^{-iM\xi}e^{-i\omega t+ik\cdot x}\psi(r)$ the equation of motion for the 
dilaton becomes
\be
\frac{1}{r^3}\partial_r(r^5\partial_r\psi)-(\mu^2M^2R^4+\frac{q^2R^4}{r^2})\psi=0,
\label{dil}
\ee
where $q^2=2M\omega+k^2$. Using AdS/CFT procedure one can compute the two
point function of the dual operator in the non-relativistic field theory.
This has essentially been done in \cite{{Son:2008ye},{Balasubramanian:2008dm}}. The result is
\be
\langle\; {\cal O}(x,t) {\cal O}(0)\;\rangle\sim t^{-\Delta} e^{-\frac{iM x^2}{2t}}=\frac{(x^2/t)^\Delta 
e^{-\frac{iM x^2}{2t}}}{x^{2\Delta}},
\ee
where $\Delta=2+\sqrt{4+\mu^2 M^2R^4}$ is the dimension of the dual operator.

As we have already anticipated in the above computations, by the scaling arguments, the solution 
of (\ref{dil}) depends only on $q^2R^4/r^2$. Therefore the UV/IR relation in the bulk and boundary
theories should be as follows \cite{Barbon:2008bg}
\be
\delta t\sim \frac{MR^4}{r^2},\;\;\;\;\;\;\;\;\delta x\sim \frac{R^2}{r}.
\ee
Actually for the relativistic case where both $x$ and $t$  scale the same, the UV/IR
relations are the same, i.e. $\delta |X|\sim R^2/r$ \cite{Susskind:1998dq}.

Since the supergravity solution (\ref{sol1}) is obtained from the type IIB D3-brane solution,
by a set of T-dualities and boosts, one expects that the resultant non-relativistic field theory,
should have gauge field and gauge symmetry. Indeed one may suspect that the obtained theory could be 
related to three
dimensional non-relativistic gauge theory which has previously been studied in the context of three dimensional 
Chern-Simons relativistic gauge theory (see for example \cite{Hagen}).
Therefore it would be interesting to explore different features of 3D non-relativistic gauge theory
by making use of the dual gravity. In particular we can evaluate
the effective potential of the external quark-anti quark potential 
via the Wilson loop computations in the context of AdS/CFT correspondence \cite{{Rey:1998ik},{Maldacena:1998im}}.

To compute the effective potential following \cite{{Rey:1998ik},{Maldacena:1998im}} we start from an
ansatz for the classical string in the supergravity solution (\ref{sol1})  which has  $Sch(d-1)$
isometry, 
\be
t=\tau,\;\;\;\;\;\;\;r=\sigma,\;\;\;\;\;\;x_1=x(\sigma),\;\;\;\;\;\;\xi={\rm constant}.
\label{OPEN}
\ee
Indeed this ansatz in the geometry (\ref{sol1}) was studied in 
\cite{Alishahiha:2003ru} where the energy of the string as a function of distance between two external
sources, $l$, was found to be
\be
E=-\frac{2\mu R^4}{\pi\alpha'}\;\frac{1}{l^2}
\ee 
This behavior may be understood from the fact that the dual theory in non-relativistic CFT with
the dynamical scaling $z=2$. 

In the rest of the paper we extend our considerations for other semi-classical strings to 
extract information about the possible operators of the dual non-relativistic 3D CFT.

\section{Semi-classical string in the background with\\ Schr\"odinger group isometry}

In this section we will study semi-classical closed strings in the non-relativistic D3-brane background.
 Since we are interested in the energy of the semi-classical string we 
need to write the non-relativistic D3-brane solution
in the ``global'' coordinates.  The corresponding solution can be found from AdS solution in the
global coordinates by making use of the Null Melvin Twist \cite{Yamada:2008if}
\bea  
ds^2&=&\left(\frac{r}{R}\right)^2\frac{1}{H}\bigg[-\left(\frac{g}{2}+\mu^2r^2  f\right)dt^2
-\frac{g}{2} d\xi^2-(1+f)dtd\xi\cr &&\cr 
&+&\frac{R^2}{4}H(d\theta^2+\cos^2\theta\;d\psi^2)\bigg] 
+\left(\frac{R}{r}\right)^2\frac{dr^2}{f}+d{\cal M}_5^2\cr &&\cr
e^{2\phi}&=&H^{-1},\;\;\;\;\;\;\;\;{\rm with}\;\;\;\;f=1+g=1+\frac{R^2}{r^2},\;\;\;\;\;\;H=1-\frac{1}{2}\mu^2 R^2.
\eea
There is also a non-zero RR four field as well as a NSNS two form. $d{\cal M}_5^2$ is the metric of internal space
whose explicit form is not important for our purpose. 
We note, however, that the detail of the internal space will become
important for other sectors of the theory. For example if we are interested in the giant magnons, the main
role is played by $d{\cal M}_5^2$. 

To proceed it is useful to make the following change of variables
\be
t\rightarrow \frac{R}{\sqrt{2}}(t+\xi),\;\;\;\;\;\xi\rightarrow \frac{R}{\sqrt{2}}(t-\xi),\;\;\;\;\;
r\rightarrow R\sinh\rho,
\ee
where upon the above metric is recast in the following form
\bea  
ds^2&=&\frac{R^2}{H}\bigg[-\left(1+\frac{\mu^2 R^2}{2}\sinh^2\rho\right)\cosh^2\rho\; dt^2
+(\sinh^2\rho-\frac{\mu^2 R^2}{8}\sinh^22\rho) d\xi^2\cr &&\cr &-&\frac{\mu^2 R^2}{4}\sinh^22\rho\; dtd\xi 
+\frac{H}{4}\sinh^2\rho(d\theta^2+\cos^2\theta\;d\psi^2)\bigg] 
+R^2d\rho^2+d{\cal M}_5^2.
\label{sol1g}
\eea

\subsection{Folded closed string}
We would like to study a solution representing a rotating 
closed string configuration which is stretched along the 
radial coordinate. In order to study this system one needs 
to write an action for this closed string.
Let us parameterize the string worldsheet by $\sigma $ and $\tau$. We can fix the re-parameterization 
invariance by a parameterization such that the time coordinate of space-time, 
$t$ to be equal to worldsheet time, i.e. $t=\tau$. 
In this gauge a closed string configuration
representing a rotating string with angular velocity 
$\omega$ on geometry (\ref{sol1g}) stretched along the radial 
coordinate is given by  
\be
t=\tau,\;\;\;\;\;\psi=\omega \tau,\;\;\;\;\;
\rho(\sigma)=\rho(\sigma+2\pi),\;\;\;\;\;\xi={\rm constant};
\label{CL}
\ee
all other coordinates are set to zero. For this solution 
the Nambu-Goto action,
\be
I=-\frac{1}{ 2\pi \alpha'}\int d\sigma^2\sqrt{-\det(G_{\mu\nu}
\partial_{a}X^{\mu}\partial_b X^{\nu})},
\label{NAMBUS}
\ee
reads
\be
I=\frac{-4R^2}{2\pi\alpha'\sqrt{H}}\int_0^{\rho_0}d\rho A^{1/2}(\rho),\;\;\;\;A(\rho)=
{(1+\frac{\mu^2R^2}{2}\sinh^2\rho)\cosh^2\rho\;{\dot t}^2- \frac{H{\dot \psi}^2}{4}\sinh^2\rho}\;,
\ee
where dot represents derivative with respect to $\tau$. 
The factor of 4 comes from the 
fact that we are dealing with a folded closed string.
Working with one fold string, the string can be divided to 
four segments. Using the periodicity condition we just need 
to perform the integral for one quarter of string
multiplied by factor 4.

To insure that the ansatz (\ref{CL}) represents a closed string we need to impose the periodicity condition
which in our case is $A(\rho)\geq 0$ for all $\rho>0$. The periodicity condition, setting ${\dot t}=1,\;
{\dot \psi}=\omega$, can be satisfied if
\be
\omega^2\geq 4 \frac{{\sqrt{2}}+{\mu R}}{{\sqrt{2}}-{\mu R}}=\omega_c^2.
\label{condition}
\ee
for which the $A(\rho)$  takes positive values for $\rho\leq \rho_-$ or $\rho\geq \rho_+$ with $\rho_-< \rho_+$, where
\be
\rho_\pm=\sinh^{-1}\left[\frac{\left((\frac{H\omega^2}{4}-1-\frac{\mu^2R^2}{2})\pm\sqrt{(\frac{H\omega^2}{4}-1-
\frac{\mu^2R^2}{2})^2-2\mu^2R^2}\right)^{1/2}}
{\mu R}\right].
\ee
An interesting feature of this semi-classical folded string is
that the closed string cannot be longer than a maximum size given
by $\rho_{max}= \sinh^{-1}[{2}^{1/4}/\sqrt{\mu R}]$ which corresponds to the length of string
whose quantum number satisfies $\omega=\omega_c$. We note that for $\rho\geq \rho_+$, even 
though $A(\rho)$ is positive, we will not get a closed string.  
This might be thought of as
the case when the periodicity condition is going to be lost and we
are dealing with open string stretched all the way to infinity.

The two conserved momenta conjugate to $t$ and $\psi$ 
are the space-time energy $E$ and spin $S$. When the periodicity condition is satisfied, using the above 
Nambu-Goto action the conserved quantities are given by
\bea
E&=&\frac{2R^2}{ \pi\alpha'\sqrt{H}}\int_0^{\rho_-}
d\rho\frac{(1+\frac{\mu^2R^2}{2}\sinh^2\rho)\cosh^2\rho}
{\sqrt{(1+\frac{\mu^2R^2}{2}\sinh^2\rho)\cosh^2\rho- \frac{H{\omega}^2}{4}\sinh^2\rho}}\;,
\cr &&\cr 
S&=&\frac{\omega R^2\sqrt{H}}{2
\pi\alpha'}\int_0^{\rho_-}d\rho\frac{\sinh^2\rho}
{\sqrt{(1+\frac{\mu^2R^2}{2}\sinh^2\rho)\cosh^2\rho- \frac{H{\omega}^2}{4}\sinh^2\rho}}\;,
\label{ES5}
\eea
From the integrals (\ref{ES5}) one can proceed to compute 
the relation between energy and spin. To do this we can use 
an approximation in which the string
is much shorter or of order of the critical value $\rho_{max}$. In other words we will consider the cases
where $\rho_-\ll \rho_{max}$  or $\rho_-\sim\rho_{max}$. Setting $\omega^2=\omega_c^2+\frac{4}{H}\eta$ the two limits
correspond to $\eta\rightarrow \infty$ and $\eta\rightarrow 0$, respectively.
Our aim will then be to find the energy $E$ as a function of spin for these two cases.

\vspace*{.3cm}

{\bf Short strings}

\vspace*{.3cm}

In this case one has $\rho_-\sim \frac{1}{\sqrt{\eta}}$
as $\eta\rightarrow\infty$. Therefore the string is much shorter than 
the radius of curvature of the geometry (\ref{sol1g}). In fact in this
limit the background space may be approximated by a flat metric 
near the center. Therefore the calculations reduce to 
spinning string in the flat space. In this limit the integrals
(\ref{ES5}) can be performed and find 
\be
E=\frac{R^2}{\alpha'\sqrt{H}}\;\frac{1}{\sqrt{\eta}},\;\;\;\;\;\;\;
S=\frac{R^2}{ 4\alpha'}\;\frac{1}{{\eta}}
\ee
so that
\be
E^2=\frac{4R^2}{H\alpha'}\; S,
\ee
which is the well-known  flat space Regge trajectory. Indeed this is what we would expect to find; 
namely going deep into the core of the space time the physics should be independent of 
general structure and the string should locally feel the flat space. This may be compared with 
AdS result where we also get the Regge trajectory, though in our case we have 
an extra factor of $1/H$ which should be the signature of the non-relativistic nature of the theory.

\vspace*{.3cm}

{\bf Near $\rho_{max}$ string}

\vspace*{.3cm}

As we have seen the closed string cannot be longer 
than a maximum size given by $\rho_{max}= \sinh^{-1}[{2}^{1/4}/\sqrt{\mu R}]$. So another limit we may 
consider is the case where string is of order of $\rho_{max}$. In this case the spin is always large 
compare with the radius of the curvature of the background geometry, i.e. $S\gg \frac{R^2}{\alpha'}$.
For $\rho_-\rightarrow \rho_{max}$ the integrals of (\ref{ES5}) yield to the following expressions
for $S$ and $E$ at leading order
\bea
E&\approx&\frac{2R^2}{\pi\alpha'}\frac{\sqrt{2}}{\mu R}\bigg{[}
\frac{1+\frac{\mu R}{\sqrt{2}}}{\sqrt{1-\frac{\mu R}{\sqrt{2}}}}\tanh^{-1}\left(\sqrt{1+\frac{\mu R}{\sqrt{2}}}
\tanh\rho_-\right)-\frac{\mu^2R^2\sinh2\rho_-}{4\sqrt{(1+\frac{\mu R}{\sqrt{2}})(2-{\mu^2R^2})}}
\cr &&\cr 
&&\;\;\;\;\;\;\;\;\;\;\;\;\;\;\;-\frac{\mu^2R^2+2\sqrt{2}\mu R+4}{4\sqrt{1-\frac{\mu^2R^2}{2}}}\rho_-
\bigg{]}+\cdots\cr &&\\
S&\approx&\frac{R^2}{\pi\alpha'}\frac{\sqrt{2}}{\mu R}\bigg{[}
\sqrt{1+\frac{\mu R}{\sqrt{2}}}\tanh^{-1}\left(\sqrt{1+\frac{\mu R}{\sqrt{2}}}
\tanh\rho_-\right)-(1+\frac{\mu R}{\sqrt{2}})\rho_-\bigg{]}+\cdots.\nonumber
\eea
Note that both of the above expressions diverge for $\rho_-\rightarrow \rho_{max}$, nevertheless for 
$\mu R\sim 1$ we find
\be
E\approx2\sqrt{\frac{\sqrt{2}+\mu R}{\sqrt{2}-\mu R}} S-\frac{R^2}{2\pi\alpha'}\;
\frac{2\sqrt{2}+\mu R}{\sqrt{2-{\mu^2R^2}}}\;{\sinh^{-1}\left(\sqrt{{\sqrt{2}}/{
{{\mu R}}}}\right)},
\ee
while for $\mu R\ll 1$ where we can expand the above expressions in terms of $\mu R$ one gets
\bea
&&E\approx\left(2+\frac{7\mu^2R^2}{64}+{\cal O}(\mu^4 R^4)\right)S
+\frac{R^2}{ \pi\alpha'}\left(1-\frac{\mu R}{8\sqrt{2}}+\frac{3\mu^2R^2}{32}
+{\cal O}(\mu^3 R^3)\right)\ln(\frac{\alpha'S}{ R^2}).\cr &&
\label{nonES}
\eea
This has to be compared with the case of $AdS_5$ geometry where we have \cite{Gubser:2002tv}
\be
E=S+\frac{R^2}{ \pi\alpha'}\ln(\frac{\alpha'}{ R^2}S)+\cdots.
\label{ES}
\ee
in which the field theory dual is a relativistic conformal gauge theory and  this behavior 
looks very similar to the logarithmic growth of anomalous dimensions of operators with spin in 
the gauge theory. Although in our case we still get the same expression as that in the
relativistic case\footnote{Note that setting $\mu R=0$ in the equation (\ref{nonES}) we recover the relativistic (\ref{ES}) up to 
numerical factors. These disagreements are due to the normalization of the $\psi$ coordinate
in metric (\ref{sol1g}). This is also the case in the short string limit.}, 
the coefficients are corrected by functions of $\mu R$, at least up to the order we are
considering. This may be understood as follows. In fact as we argued the dual non-relativistic theory
contains a gauge field inherited from the 4D ${\cal N}=4$ SYM theory. So, the 3D theory may be
studied by a small deformation of the 4D where the deformation parameter is $\mu R$. Therefore
for $\mu R\ll 1$ we still have the same operators as those in 4D 
though the 
anomalous dimension of the corresponding operators get corrected due to the deformation, 
bringing us to the non-relativistic field theory.
It would be interesting to see if such a behavior can be obtained from non-relativistic gauge 
theory as well.

\subsection{Circular pulsating string}

Another semi-classical string we would like to study is a circular pulsating string first studied in
\cite{Minahan} in the AdS geometry. This is a string which wrapped around a angular coordinate and
pulsates in radial direction. More precisely consider a circular pulsating closed string which is 
wrapped $m$ times around the $\psi$ direction. The corresponding string configuration is given by
\be
t=\tau,\;\;\;\;\;\;\;\rho=\rho(t),\;\;\;\;\;\;\;\psi=m\sigma,\;\;\;\;\;\;\xi={\rm constant}.
\ee 
The other coordinates are set to zero. The Nambu-Goto action for this configuration in the geometry (\ref{sol1g})
reads
\be
S=-\frac{mR^2}{4\alpha'\sqrt{H}}\int dt\;(1+\frac{\mu^2R^2}{2}\sinh^2\rho)^{1/2}\;\sinh2\rho\;
\sqrt{1-\frac{H{\dot \rho}^2}{(1+\frac{\mu^2R^2}{2}\sinh^2\rho)\cosh^2\rho}},
\ee
where dot represents derivative with respect to $t$. It is useful to make the following 
change of variable
\be
\eta=\int \sqrt{\frac{H}{(1+\frac{\mu^2R^2}{2}\sinh^2\rho)\cosh^2\rho}}\;d\rho,
\ee
by which the above action can be recast to the following form
\be
S=-\int dt\;\; g(\eta)\;\sqrt{1-{\dot \eta}^2},
\ee
where $g(\eta)=\frac{mR^2}{4\alpha'\sqrt{H}}(1+\frac{\mu^2R^2}{2}\sinh^2\rho)^{1/2}\;\sinh2\rho$.
The associated Hamiltoninan with the above action is
given by
\be
H=\sqrt{\Pi^2+g(\eta)^2}
\ee
with $\Pi$ being the canonical momentum. Note that the $H^2$ may be considered as a
one dimensional quantum mechanical system with the potential
\be
V(\eta)=g(\eta)^2=\left(\frac{mR^2}{4\alpha'\sqrt{H}}\right)^2(1+\frac{\mu^2R^2}{2}\sinh^2\rho)\;\sinh^22\rho.
\ee
Therefore following \cite{Minahan} we can use the Bohr-Sommerfeld analysis for the quantization of the states. The 
quantization condition is 
\be
(n+\frac{1}{2})\pi=\int_{\eta_1}^{\eta_2} d\eta\;\;\sqrt{E^2-V(\eta)}
\ee
where $\eta_{1,2}$ are the turning points. It is useful to return to the original coordinate $\rho$ 
in which the above quantization condition becomes
\be
(n+\frac{1}{2})\pi=E\sqrt{H}\int d\rho\;\sqrt{\frac{{1-\frac{1}{B^2}
(1+\frac{\mu^2R^2}{2}\sinh^2\rho)\sinh^22\rho}}
{ (1+\frac{\mu^2R^2}{2}\sinh^2\rho)\cosh^2\rho}},
\ee
where $B^2=\frac{4\alpha'E\sqrt{H}}{mR^2}$. To perform the integral  we follow the procedure of
\cite{Minahan, Alishahiha} decomposing the integral into two 
parts
\bea
(n+\frac{1}{2})\pi&=&E\sqrt{H}\bigg[-
\int_0^{\rho_0} d\rho\;
\frac{1-\sqrt{{1-\frac{1}{B^2}
(1+\frac{\mu^2R^2}{2}\sinh^2\rho)\sinh^22\rho}}}
{ \sqrt{(1+\frac{\mu^2R^2}{2}\sinh^2\rho)\cosh^2\rho}}\cr &&\cr
&&\;\;\;\;\;\;\;\;\;\;\;\;\;+\int_0^{\rho_0} \frac{d\rho}
{\sqrt{ (1+\frac{\mu^2R^2}{2}\sinh^2\rho)\cosh^2\rho}}\;\bigg],
\label{int1}
\eea
where $\rho_0$ is the turning point in the original coordinates. 
For the large $B$ the first integral
in (\ref{int1}) becomes
\be
\int_0^{\rho_0} d\rho\;
\frac{1-\sqrt{{1-\frac{1}{B^2}
(1+\frac{\mu^2R^2}{2}\sinh^2\rho)\sinh^22\rho}}}
{ \sqrt{(1+\frac{\mu^2R^2}{2}\sinh^2\rho)\cosh^2\rho}}
\approx \frac{2B^{-2/3}}{(\sqrt{2}\mu R)^{1/3}}\;
\left(-\frac{1}{2}+\frac{3^{3/2}\Gamma\left(\frac{2}{3}\right)^{3}}{2^{8/3}\pi}\right),
\ee
while for the second integral one finds
\be
\int_0^{\rho_0} \frac{d\rho}
{\sqrt{ (1+\frac{\mu^2R^2}{2}\sinh^2\rho)\cosh^2\rho}}
\approx \frac{2B^{-2/3}}{(\sqrt{2}\mu R)^{1/3}}\;
\left(\frac{2\pi^2}{9\Gamma\left(\frac{2}{3}\right)^{3}} B^{2/3}-\frac{1}{2}\right).
\ee
Thus altogether we get
\be
(n+\frac{1}{2})\pi\approx \frac{4\pi^2}{9\Gamma\left(\frac{2}{3}\right)^3}\;\frac{\sqrt{H}}
{(\sqrt{2}\mu R)^{1/3}}\;E
-\frac{3^{3/2}\Gamma\left(\frac{2}{3}\right)^3}{8\pi}\;\left(\frac{\sqrt{H}}{\sqrt{2}\mu{\alpha'}^2}\right)^{1/3}\;
Rm^{2/3} E^{1/3},
\label{mn}
\ee
which can be inverted to find energy as a function of $n$
\be
E\approx
\alpha n+\beta(m^2n)^{1/3},
\ee
where $\alpha$ and $\beta$ are two constants given in terms of $\mu,R$ and $\alpha'$ which can be 
read from equation (\ref{mn}), though whose explicit forms are not important for our consideration.
This has to be compared with that in AdS geometry where it was found that the $E-n$ grows as $n^{1/2}$
\cite{Minahan}. 
It would be interesting to find the dual operator in the non-relativistic gauge theory.

\section{Open string and drag force}

So far we have considered closed strings in the geometry with Schr\"odinger isometry. 
As we have already mentioned, in section two, open string can also be used to explore different 
features of the model using the supergravity dual. It may be used to obtain, for example, the effective potential
of quark-anti quark system. In this section we would like to study a non-relativistic 
quark moving through a hot plasma by making use of the gravity description of the system.
In other words, following \cite{Gubser} we would like to study 
the drag force for a quark in a non-relativistic field theory.

To do so, one first needs the gravity dual of the non-relativistic field theory
at finite temperature. In the context of AdS/CFT duality we know that heating up the dual field theory
generically corresponds to adding a black hole in the bulk geometry. Therefore
 we need to find the supergravity solution corresponding to the 
black hole in the geometry (\ref{sol1}). The relevant supergravity solution for
our studies is given in  \cite{Mazzucato:2008tr} (see also \cite{Kovtun:2008qy}) 
\bea  
ds^2&=&\left(\frac{r}{R}\right)^2\frac{1}{H}\left[-\left(\frac{g}{2}+{\mu^2 r^2f}\right)dt^2
-\frac{g}{2} d\xi^2+(1+f)dtd\xi+H dx_i^2\right]\cr &&\cr
&+&\left(\frac{R}{r}\right)^2\left[\frac{dr^2}{f}+r^2\left(\frac{(d\chi+A)^2}{H}+ds_P^2\right)\right],\cr &&\cr
e^{2\phi}&=&H^{-1} 
\label{sol2}
\eea
where 
\be
g=-\left(\frac{r_H}{r}\right)^4,\;\;\;\;\;\;\;\;\;H=1-\frac{\mu^2r^2}{2}g,\;\;\;\;\;\;\;f=1+g
\ee
There is also a non-zero RR 4-form as well as a NSNS 2-form (see \cite{Mazzucato:2008tr}). 

To proceed we start from an ansatz for the open string representing an external moving source in the 
dual field theory.
To write the open string ansatz it is useful to make the 
following change of variables
\be
t\rightarrow \frac{1}{\sqrt{2}} (\xi-t),\;\;\;\;\;\;\;\;\xi\rightarrow \frac{1}{\sqrt{2}}(\xi+t),
\ee
in which the above metric reads
\bea  
ds^2&=&\left(\frac{r}{R}\right)^2\frac{1}{H}\left[-(1+\frac{1}{2}\mu^2 r^2)fdt^2
+(1-\frac{1}{2}\mu^2 r^2 f) d\xi^2+\mu^2r^2fdtd\xi+H dx_i^2\right]\cr &&\cr
&+&\left(\frac{R}{r}\right)^2\left[\frac{dr^2}{f}+r^2\left(\frac{(d\chi+A)^2}{H}+ds_P^2\right)\right].
\label{sol3}
\eea
In this notation our ansatz is given by
\be
t=\tau,\;\;\;\;\;\;r=\sigma,\;\;\;\;\;\;x_1=vt+x(r),\;\;\;\;\;\;\xi={\rm constant}.
\label{drag}
\ee
Re-writing the relevant part of the metric in the following form
\be
ds^2=g_{tt}dt^2+g_{xx}dx_1^2+g_{rr}dr^2
\ee
the Nambu-Goto action becomes \cite{Herzog}
\be
S=-\frac{1}{ 2\pi \alpha^{'}}\int dt dr \sqrt{-( g_{tt}g_{rr}+g_{tt}g_{xx} {x'}^{2}+g_{xx}g_{rr}v^{2})}
\ee
where prime represents derivative with respect to $r$. Since the metric components are $t$ independent,
the above action may be treated as a one dimensional mechanical system whose momentum is the
constant of motion
\be
\frac{-g_{tt}g_{xx} x'}{\sqrt{-( g_{tt}g_{rr}+g_{tt}g_{xx} {x}'^{2}+g_{xx}g_{rr}v^{2})}}
=c=-2\pi\alpha' \pi_x={\rm constant},
\ee
which can be solved for $x'$ leading to
\begin{eqnarray}
{x'}^{2}=4\pi^2{\alpha'}^2\pi_x^{2}\bigg{(}\frac{g_{rr}(-g_{tt}-g_{xx}v^{2})}{g_{xx}g_{tt}(g_{xx}g_{tt}
+4\pi^2{\alpha'}^2\pi_x^{2})}\bigg{)}.
\label{a2}
\end{eqnarray}
In terms of the constant $\pi_x$ one has \cite{Herzog} 
\be
\frac{dE}{dt}={\pi_xv} ,\;\;\;\;\;\;\;\;\frac{dP}{dt}=\pi_x.
\label{EP}
\ee
where $E$ and $P$ are energy and momentum the open string gain from through its end point.
To find $c$ we note that the equation (\ref{a2}) physically make sense  if the numerator and denominator 
vanish at the same point \cite{Gubser}. Setting the numerator of (\ref{a2}) to zero, for the supergravity solution
(\ref{sol3}), one finds
\begin{eqnarray}
\frac{1}{2}\mu^{2}r^{6}_0+(1-v^{2})r^{4}_0-\frac{1}{2}\mu^{2}r_{H}^{4}r^{2}_0(1+v^{2})-r_{H}^{4}=0,
\label{r0}
\end{eqnarray}
which can be solved for $r_0$. Plugging the solution $r_0$ in the denominator one arrives at
\be
\pi_x=-\frac{v}{2\pi\alpha'}g_{xx}|_{r_0} .
\label{mom}
\ee
From (\ref{r0}) we see that setting $\mu=0$ the velocity changes from $v=0$ to $v=1$ as $r_0$ varies
from $r_H$ to infinity, as expected for relativistic field theory. On the other hand for $\mu\neq 0$
where the dual theory is supposed to be non-relativistic we observe that as we are varying the  $r_0$
from $r_H$ to infinity, the velocity takes its value from zero to infinity. This is in fact due to 
the non-relativistic property of the dual field theory.

To proceed we need to solve the equation (\ref{r0}) to find $r_0$ in terms of velocity. 
Then using the expression for 
constant conjugate momentum in terms of the metric components presented in (\ref{mom}), one may read, 
for example, the drag force from (\ref{EP}). To proceed we will consider two different limits
depending on whether the velocity is small or large. In these limits the drag force becomes 
\be
\frac{dP}{dt}\approx
\left\{
\ba{lll}
&-\frac{v}{2\pi\alpha'}\;\frac{r_H^2}{R^2}\; (1+\frac{1}{2}v^2)\;\;\;\;\;\;\;\;\;\;\;\;&{\rm for}\;\;\;\;\;v\ll 1,
\cr &&\cr
 &-\frac{v}{2\pi\alpha'}\;\frac{2}{\mu^2 R^2} (v^2+{\mu^4r_H^4-4})\;\;\;\;\;\;&{\rm for}\;\;\;\;\;v\gg 1.
 \ea
 \right.
 \ee
We recognize the first one as the non-relativistic limit of that found in 
\cite{Gubser} for the relativistic field theory when $v\ll 1$. The second case is  just because of 
the non-relativistic nature of the dual field theory.  
 
Now consider a single non-relativistic particle with momentum $P$ and mass $M$, then we have 
$P=Mv$. It is useful to formally rewrite the above expression for the drag force in terms of $P$. Then 
we can perform the integral yielding
 \be
P(t)\approx
\left\{
\ba{lll}
&P_0\ e^{-\frac{\pi R^2 T^2}{4\alpha'M}\;t}\;\;\;\;\;\;\;\;\;\;\;\;&{\rm for}\;\;\;\;\;P_0\ll M,
\cr &&\cr
 &\left(\frac{1}{P^2_0}+\frac{2t}{\pi\alpha'\mu^2 R^2M^3}\right)^{-1/2}
 \;\;\;\;\;\;&{\rm for}\;\;\;\;\;P_0\gg M.
 \ea
 \right.
 \label{pp}
 \ee
In the above expression we have set $r_H^2=\frac{\pi^2 R^4}{2} T^2$ \cite{Mazzucato:2008tr}. 
Similarly one finds
  \be
E(t)\approx
\left\{
\ba{lll}
&E_0\ e^{-\frac{\pi R^2 T^2}{2\alpha'M}\;t}\;\;\;\;\;\;\;\;\;\;\;\;&{\rm for}\;\;\;\;\;E_0\ll M/2,
\cr &&\cr
 &\left(\frac{1}{E_0}+\frac{4 t}{\pi \alpha'\mu^2 R^2M^2}\right)^{-1}
 \;\;\;\;\;\;&{\rm for}\;\;\;\;\;E_0\gg M/2.
 \ea
 \right.
 \ee
This means that a particle with energy much less than its mass will lose its energy 
exponentially with time, while for a particle with kinetic energy much more than its
mass, the energy is lost as $t^{-1}$. Another interesting feature of the model is that
for the slowly moving particles the relaxation time, $t_0 =\frac{2\alpha'M}{\pi R^2 T^2}$,
 depends inversely on temperature, whereas for the fast moving particles it is 
temperature independent, $t_0={\pi \alpha'\mu^2 R^2M^2}/4$. On the other hand
in the first case the relaxation time is $\mu$ independent, while in the second case
it proportional to $\mu^2$. This shows that even at zero temperature a non-relativistic 
particle will lose its energy. In the next section we explore this point in more details.

\section{Speed limit and drag force} 
 
To study a quark moving through a hot plasma, Gubser \cite{Gubser} has considered a
moving open string in a geometry with horizon where the radius of the horizon is 
related to the temperature of the dual gauge theory. Although having the horizon is 
important to deal with the gauge theory at finite temperature, as far as the gravity 
computations are concerned we are free to redo the computations for a geometry without 
horizon. 
Essentially what one needs to do is Wilosn loop computations in the context of 
AdS/CFT correspondence where the string is moving as well. 

Let us first consider a moving open string given by (\ref{drag}) in the $AdS_5\times S^5$ geometry
parametrized as follows
\be
ds^2=\left(\frac{r}{R}\right)^2(-dt^2+dx_1^2+dx_2^2+dx_3^2)+\left(\frac{R}{r}\right)^2dr^2+R^2d\Omega_5^2.
\ee
Using the procedure of the previous section one gets
\be
{x'}^2=(2\pi \alpha')^2\pi_x^2\frac{1-v^2}{\left(\frac{r}{R}\right)^4\left(\left(\frac{r}{R}\right)^4-
(2\pi \alpha')^2\pi_x^2\right)},
\ee
which is well behaved if $v<1$ representing the fact that the dual theory is relativistic and therefore
there is a bound for the velocity. Moreover we observe that the constant conjugate momentum, $\pi_x$, is 
independent of $v$. 

Now consider the following ansatz for the open string moving in the background (\ref{sol1})
\be
t=\tau,\;\;\;\;\;r=\sigma,\;\;\;\;\;x_1=vt+x(r),\;\;\;\;\;\xi={\rm constant}. 
\ee
In this case one finds
\be
{x'}^2=(2\pi \alpha')^2\pi_x^2\frac{\mu^2r^2-v^2}{\left(\frac{r}{R}\right)^4\mu^2r^2
\left(\left(\frac{r}{R}\right)^4\mu^2r^2-
(2\pi \alpha')^2\pi_x^2\right)}.
\ee
We observe that in this case there is no bound on the velocity and it can change from zero to
infinity. This is indeed the reflection of the fact that dual theory is non-relativistic.
To avoid the imaginary solution one arrives at
\be
\pi_x=-\frac{v}{2\pi\alpha'}\;\frac{v^2}{\mu^2 R^2}.
\ee
For a particle with mass $M$ and momentum $P=Mv$ the drag force reads
\be
\frac{dP}{dt}=-\frac{1}{2\pi\alpha'M^3\mu^2 R^2}\;P^3,
\ee
which yields to
\be
P=\left(\frac{1}{P_0^2}+\frac{t}{\pi \alpha' M^3 \mu^2 R^2}\right)^{-\frac{1}{2}}.
\ee
This means that in this case even though the system is at zero temperature the moving
particle losses its energy and the relaxation time is given in terms of $\mu^2$. Whereas in the
hot plasma it is controlled by temperature. This may be be related to the fact that $\mu$ in the
dual non-relativistic field theory could be interpreted as the chemical potential.
  
The drag force calculations 
can be generalized to other backgrounds obtained from Null Melvin Twist 
procedure of $Dp$-brane for $p\leq 4$. The relevant solutions are given by \cite{Mazzucato:2008tr}
\bea\label{dp}  
ds^2&=&\left(\frac{r}{R}\right)^{\frac{7-p}{2}}\frac{1}{H}\left[-(1+\frac{1}{2}\mu^2 r^2)fdt^2
+(1-\frac{1}{2}\mu^2 r^2 f) d\xi^2+\mu^2r^2fdtd\xi+H dx_i^2\right]\cr &&\cr
&+&\left(\frac{R}{r}\right)^{\frac{p+1}{2}}\left[\frac{dr^2}{f}+r^2\left(\frac{(d\chi+A)^2}{H}+ds_P^2\right)\right],
\eea
where $f=1-(r_H/r)^{7-p}$ and $H=1+\mu^2 r_H^{7-p}/2r^{5-p}$. Going through the computations of section
four, for slowly moving particles, we find
\be
P=P_0\;e^{-\frac{t}{t_0}},\;\;\;\;\;\;{\rm with}\;\;\;\;\;\;\;t_0=\left(\frac{(7-p)^4}
{2^{p+1}\pi^{p-1}}\right)^{\frac{1}{5-p}}\;\frac{\alpha' M}{(RT)^{\frac{4}{5-p}}},
\ee
while for fast moving particles we get a universal result given by that in equation (\ref{pp}).
Therefore all the models parametrized by $p\leq 4$ exhibit the same non-relativistic behavior, though 
in the case of slowly moving particles the characteristic nature of energy lost is fixed by 
different power of the temperature. i.e.  $t_0\sim (RT)^{\frac{4}{p-5}}$.

\section{Discussions and conclusions  }

In this paper we have studied a number of features of non-relativistic CFT 
by making use of the supergravity solution in type IIB string theory. Although we have mainly considered  
3 dimensional CFT whose gravity dual can be obtained from D3-brane
using TsT duality, we would expect that the general features we explored in this paper 
can be applied for other dimensions too.

Since the world volume theory of D3 brane is a
supersymmetric gauge theory, and taking into account that supergravity solution (\ref{sol1}) is
obtained from D3-brane by  the Null Melvin Twist procedure, we would expect that the resultant theory 
still contains a gauge field. Of course the procedure will
reduce the number of supersymmetries as well as the space time symmetry. Indeed 
for the case where the internal space is a sphere the amount of supercharge preserved by the
background are eight. The space time symmetry will also reduce to Schr\"odinger symmetry.
Therefore we expect that the field theory dual to the type IIB on supergravity solution (\ref{sol1}) to be non-relativistic super conformal gauge theory in three dimensions.
 
Since the supergravity dual can be embedded in type IIB string theory it is natural to study
semi-classical string in this background to explore some properties of the dual non-relativistic
superconformal gauge theory. In particular we have seen that the effective potential of the
external objects is proportional to $l^{-2}$ as expected for a non-relativistic CFT. 
One may find the effective potential of quark-anti quark as a function of
distance $l$ for arbitrary $p$ where the corresponding supergravity solutions are given by
(\ref{dp}) with $r_H=0$. This has been done in \cite{Alishahiha:2003ru} where the authors argued that
the interaction between external objects is due to their lightlike dipole moments given by
$\mu$ 
\be
E\sim -\frac{\mu}{\alpha'}\;\left(\frac{R^4}{l^2}\right)^{\frac{2}{5-p}}.
\ee
For $p=3$ which we have mainly considered in this paper, we have interpreted the effective 
potential due to non-relativistic CFT given by dynamical scaling $t\rightarrow \lambda^2 t$
and $x\rightarrow \lambda x$. For other cases there is no such a clear interpretation. 

We have also considered folded rotating closed strings in the 
geometry with isometry of Schr\"odinger group where we have shown that the anomalous
dimension of the corresponding dual operator exhibits logarithmic corrections similar to 
that in the relativistic gauge theory for $\mu R\ll 1$. 
It would be interesting to find such a behavior directly from non-relativistic gauge theory.

We have also studied circular pulsating closed strings where we have observed that although in the leading 
order the anomalous dimension is proportional to the winding number of the string, $\Delta\sim n$, at subleading 
order it goes as $n^{1/3}$. It is worth noting that in the case of four dimensional relativistic gauge theory 
the subleading correction grows as $\Delta-n\sim n^{1/2}$. This should be taken as the effect of non-relativistic
nature of the theory.

It is worth noting that whenever we have deviations from the relativistic field theory or in the
gravity side from $AdS_5$ gravity, the deviations are controlled by the dimensionless parameter
$\mu R$. Therefore assuming $\mu R\ll 1$ this parameter can be thought of as the expansion parameter 
by which we can study the non-relativistic three dimensional conformal gauge theory as a 
perturbation of four dimensional ${\cal N}=4$ superconformal gauge theory.
In particular since the non-relativistic CFT we have been considering may be obtained by reduction
(contract) from ${\cal N}=4$ 4D theory, we might suspect that the AdS/CFT dictionary, in some extend,
works the same as before. If correct, the operator dual to dilaton would be ${\rm Tr} F^2$, though in this case
the anomalous dimension of the operator gets higher loop corrections, such that summing up all the
loops we get $\Delta=2+\sqrt{2+n^2 \mu^2 R^2}$. Actually 
this is the expression we have given in section two for the case of $M=\frac{n}{R}$ where from gravity 
point of view $M$ is the momentum of the dilaton in the light like direction $\xi$. 
By $F^2$ we mean the reduction of four dimensional $F^2$ to three dimensions.

On the other hand since the three dimensional theory is a supersymmetric theory we expect to have 
scalars in the model which might be identified with the coordinates in which the isometry of internal 
space acts on. Therefore following the general  philosophy of \cite{Gubser:2002tv} one might expect that 
the operators dual to the folded rotating closed string with spin $S$ have the following schematic form
\be
{\cal O}_S\sim {\rm Tr} X \nabla^S X,
\ee
where $X$ is the three dimensional scalar filed. At leading order the anomalous dimension should 
be $\Delta\sim S$. But as we have seen the anomalous 
dimension of the operator gets corrections and the corrections depend
on $\mu R$ which controls the non-relativistic effects. It would be interesting to study 
these operators from non-relativistic gauge theory point of view.

We have also considered non-relativistic three dimensional CFT at finite temperature.
In particular we have considered an open string moving in the background created by a
black hole geometry in (\ref{sol1}). In the dual picture
this means that we are dealing with a quark moving through the hot plasma. Following \cite{Gubser}
we have evaluated the drag force for this case too. The first observation we have made is that
 there is no speed limit 
in this case, pointing toward the fact that the theory is non-relativistic\footnote{The main motivation of 
the paper \cite{Alishahiha:2003ru} where the metric (\ref{sol1}) was first presented was to study the
non-local features of the dual theory. In light of the recent studies as well as our observation in this paper
we see that the dual theory is indeed non-local due to is non-relativistic nature.}. Moreover depending on the initial energy
of the moving quark, it loses the energy either exponentially $e^{-t/t_0}$ or as $(t_0/t)$.
We have also observed that the relaxation time for slowly moving particles depends on the temperature and 
is independent of $\mu$,
though for fast moving particles it only depends on $\mu$.

The drag force like computations may also be done for the non-relativistic field theory even at zero
temperature. Doing so, one finds that unlike the AdS case where we only get a casual speed limit,
in the non-relativistic case one arrives at non-trivial results, though in comparison with finite
temperature system, the physics is controlled by $\mu$. It would be interesting to understand
this observation from non-relativistic gauge theory dual.

\vspace*{1cm}

{\bf Acknowledgments}

 We would like to thank Farhad Ardalan for useful discussions and comments. We would
 also like to thanks Shahin Rouhani for discussions on the AdS/non-relativistic CFT
 duality. This work is supported in 
part by Iranian TWAS chapter at ISMO.

\end{document}